\newcommand{\be}{\begin{equation}}
\newcommand{\ee}{\end{equation}}
\newcommand{\bea}{\begin{eqnarray}}
\newcommand{\eea}{\end{eqnarray}}
\newcommand{\al}{\alpha}
\newcommand{\pa}{\partial}
\newcommand{\ga}{\gamma}
\newcommand{\bet}{\beta}
\newcommand{\vrho}{\varrho}
\newcommand{\ka}{\kappa}
\newcommand{\de}{\delta}
\newcommand{\vphi}{\varphi}
\newcommand{\ta}{\theta}
\newcommand{\vth}{\vartheta}
\newcommand{\Th}{\Theta}
\newcommand{\rar}{\rightarrow}
\newcommand{\zp}{z^{+}}
\newcommand{\lb}{\left[}
\newcommand{\rb}{\right]}
\newcommand{\zm}{z^{-}}
\newcommand{\pam}{\partial_{-}}
\newcommand{\pap}{\partial_{+}}
\newcommand{\non}{\nonumber}
\newcommand{\pr}{{\rm P\!}_{-}}
\newcommand{\tr}{{\rm tr}}
\newcommand{\ra}{\rangle}
\newcommand{\la}{\langle}
\newcommand{\gp}{{\cal G}_{+}}
\newcommand{\gn}{{\cal G}_{0}}
\newcommand{\gm}{{\cal G}_{-}}
\newcommand{\emt}{energy momentum tensor}
\newcommand{\ba}{\begin{array}}
\newcommand{\ea}{\end{array}}

\documentstyle[12pt]{article}

\begin{document}
\vspace{-1mm}
\begin{flushright} G\"{o}teborg ITP 96-2 \\
{\tt hep-th/9603040}
\end{flushright}
\vspace{1mm}
\begin{center}{\bf\Large\sf Generalized Toda Theories from
 WZNW Reduction}
\end{center}

\begin{center}{{\bf\large Niclas Wyllard{\normalsize
 \footnote{wyllard@fy.chalmers.se}}\vspace{5mm}}\\{\em
 Institute of Theoretical Physics, S-412 96 G\"{o}teborg,
 Sweden}} 
\end{center}

\begin{abstract}
We reconsider the, by Brink and Vasiliev, recently
 proposed generalized
 Toda field theories using the framework of WZNW$\rightarrow$Toda
 reduction. The reduced theory has a gauge symmetry which can be
 fixed in various ways. We discuss some different gauge choices.
 In particular we study the ${\cal W}$ algebra associated with the
 generalized model in some different realizations,
 corresponding to
 different gauge choices. We also investigate the mapping
 between the
 Toda field and a free field and show the relation between the
 ${\cal W}$ algebra generators expressed in terms of the two
 different fields. All results apply also to the case of ordinary
 Toda theories.
\end{abstract}

\section{Introduction}
Recently Brink and Vasiliev \cite{Brink} proposed a model
 generalizing
 the Toda (field) theories based on the simple Lie algebras
 $A_{N}$
 (the cases of $B_{N}$ and $C_{N}$ were also treated). The model
 involves a continuous parameter, such that when this parameter
 takes certain discrete values the model reduces to the ordinary
 $A_{N}$ Toda theories. In this sense the model is a universal
 theory for the  $A_{N}$, $B_{N}$, and $C_{N}$ (Toda) theories.
 The
 approach used by Brink and Vasiliev is based  on a field $\phi$
 which
 is taken to depend not only on the two dimensional space-time
 coordinates, but also on an $sl_{2}$ algebra generator, $T^{0}$.
 The field can be expanded as
 $\phi(T^{0},z) = \sum_{n}\vphi_{n}(z)h_{n}(T^{0})$,
 where $h_{n}$ is a certain $n$th order polynomial in $T^{0}$.
 It is possible to define a trace operation
 \cite{Vasiliev89} \cite{Vasiliev91} for the $h_{n}$'s such
 that $\tr(h_{n}h_{m})\propto \de_{nm}$. Using this trace
 operation, an action for the model can be constructed. In
 this paper we will reconsider this model from a different
 perspective. We will confine ourselves to the study of the
 classical
 model.

One of the interesting features of the Toda theories, as was
 first
 shown by Bilal and Gervais \cite{Bilal1} \cite{Bilal2} 
\cite{Bilal3},
 is the fact that the Toda theories possess a ${\cal W}$ 
symmetry.
 The generalized infinite-dimensional model carries a 
${\cal W}$
 symmetry algebra generalizing the ${\cal W}_{N}$ algebra, 
in
 the sense that the ${\cal W}$ algebra reduces to the
 ${\cal W}_{N}$
 algebra, for a particular choice of the continuous
 parameter in
 the model. The ${\cal W}$ algebra is  a universal algebra 
for the
 ${\cal W}_{N}$ algebras. In \cite{Brink} two different 
realizations
 of the general ${\cal W}$ algebra were given. In the 
first realization 
the ${\cal W}$ currents were expressed in terms of the
 field $\phi$. 
The $i$th generator, $W_{i}$, was shown to be the trace
 of an $i$th
 order polynomial in $\phi$. The ${\cal W}$ algebra
 is the Poisson
 bracket algebra of the $W_{i}$'s using the canonical 
commutation
 relations 
of the $\phi$-field. In the other realization of the ${\cal W}$ 
algebra a field $\mu$, depending on (besides the
 space-time coordinates)
 the $sl_{2}$ generator $T^{-}$, was introduced. It was 
argued that
 the generators of the ${\cal W}$ algebra were equal to 
the components
 of the field $\mu$. The $\mu$-field bracket used for calculating
 commutators of the ${\cal W}$ algebra, however, is not of the 
standard Darboux form; rather, as we will see in section
 \ref{diff},
 it is a Dirac bracket.  
     
It has been shown \cite{Forgacs} \cite{ORannphys} \cite{ORphysrep}
 that the Toda theories can be obtained from a WZNW model by a 
(hamiltonian) reduction. The constraints imposed on the WZNW
 model
 are first class which means that we have gauge invariance in
 the theory. In this paper we will apply the methods given in 
\cite{ORphysrep} to the generalization proposed by Brink and
 Vasiliev.

In sections \ref{wzw} and \ref{sgau} we will review the basic
 facts 
about WZNW$\rar$Toda reduction, following \cite{ORphysrep}. In 
section \ref{alg} we will describe the work of Brink and Vasiliev,
 with emphasis on the infinite-dimensional algebra underlying
 their
 work.  The ${\cal W}$ algebra appears naturally in the WZNW
 reduction
 approach as a Dirac bracket algebra. Different gauge choices
 lead to
 different (isomorphic) realizations of the $\cal W$ algebra. In
 section \ref{diff} we will discuss two particular gauge choices, 
these will be shown to lead to the two realizations of the
 ${\cal W}$
 algebra used by Brink and Vasiliev. We will obtain a natural
 explanation of the bracket for the $\mu$-field  as the Dirac
 bracket in one of the two gauges. Finally, in section \ref{sgov}
 we will examine the connection between the two gauge
 choices mentioned
 above; this will lead us to the governing equation of ref.
 \cite{Brink}.
This equation connects the ${\cal W}$ algebra generators in the 
two realizations (gauge choices). We will also touch upon the 
subject of B\"{a}cklund transformations. We end with a
 short discussion.

\section{Review of WZNW$\rar$Toda Reduction} \label{wzw}
Some years ago it was realized that the Toda theories could be 
obtained from a WZNW model by a (hamiltonian) reduction
 \cite{Forgacs} \cite{ORannphys} \cite{ORphysrep}. We will
 here briefly review some aspects of this development in a 
way which can easily be extended to the infinite-dimensional
 generalization which we will discuss in later sections. 
The WZNW action is \cite{Witten84} \cite{Novikov} 
\cite{Wess}\footnote{Signature of 
metric, $\eta_{\mu\nu} = {\rm diag}(1,-1)$;
 level $k=-4\pi\ka$.}
\be
	S_{{\rm W}}(g)=\frac{\ka}{2}\int\eta^{\mu\nu}
\tr(g^{-1}\pa_{\mu}g)(g^{-1}\pa_{\nu}g)d^{2}z - \frac{\ka}{3}
\int_{B_{3}}\tr(g^{-1}dg)^{3}\,,
\ee
where as usual $B_{3}$ is a three-dimensional manifold whose
 boundary is space-time. The field $g(z)$ is taken to be valued
 in a connected Lie group, whose associated Lie algebra,
 ${\cal G}$,
 is assumed to be simple and maximally non-compact. The
 maximally
 non-compact real form of a complex simple Lie algebra is the real
 algebra obtained by choosing the Cartan-Weyl basis in which all 
the structure constants are real numbers; the generators in this 
basis are then taken to span a real algebra, which is the
 maximally
 non-compact one. As an example, the maximally non-compact
 form of
 $A_{N}$ is $sl({N+1},{\bf R})$. Among the various $sl_{2}$
 subalgebras
 of ${\cal G}$ we will in particular be interested in
 the so called 
principal embedding\footnote{Also known as the maximal
 embedding.}. 
This special embedding will be described in more detail later. The 
details will be important in the derivation of the Toda theory. We 
denote the generators of this particular  embedded $sl_{2}$ by 
$T^{0}$, $T^{\pm}$; they satisfy the commutation relations
\bea
	[T^{0},T^{\pm}] = \pm T^{\pm} &,& [T^{-},T^{+}]=2T^{0}\,. 
\label{comm}
\eea
$T^{0}$ defines an integral gradation of the algebra by the
 eigenvalues
 of $ad_{T^{0}} = \lb T^{0}, \cdot\rb$. We may split the
 algebra into
 eigenspaces of this operator. Let ${\cal G}_{m}$ be the 
eigenspace 
corresponding to the eigenvalue $m$; for future reference we
 introduce 
the notation
\bea
	\gm = \bigoplus_{m<0} {\cal G}_{m}&,&\gp = \bigoplus_{m>0}
 {\cal G}_{m}\,, 
\eea
where the direct sum is to be considered as a sum of vector
 spaces.
 We denote the basis elements of ${\cal G}$ by $E^{s}_{n}$,
 where $s$ 
is the eigenvalue of $ad_{T^{0}}$, and $n$ is an additional label 
needed to completely specify the element. The subset consisting of 
basis elements $E_{n}^{s}$ with $s>0$ ($s<0$) form a basis of
 $\gp$ ($\gm$). The set of elements $h_{n}=E^{0}_{n}$ form a basis
 of $\gn$. Elements which lie in 
$\ker ad_{T^{+}}$ ($\ker ad_{T^{-}}$)\footnote{The kernel of an
 operator $A$, $\ker A$,
 is defined as the set of all $x$ such that $Ax=0$.}
will be called highest (lowest) weight.
With the exception of (the maximally non-compact form of) $D_{2n}$,
 it can be shown that there is no more than one  highest (lowest)
 weight elements for each fixed $T^{0}$ grade.
The basis elements can in all cases be chosen to satisfy
\be
 	\tr(E^{s}_{n}E^{r}_{m}) = \de_{s+r}\de_{nm}\,, \label{norm}
\ee
which means that only elements which have zero grade with 
respect to $ad_{T^{0}}$ can have a non-zero trace. The WZNW 
currents are $J_{+}=\ka\pap gg^{-1}$ and $J_{-}=-\ka g^{-1}\pam g$;
 the equation of motion is, $\pa_{+} J_{-}=0$, or equivalently
 $\pa_{-} J_{+}=0$.\footnote{Conventions: $z^{\pm} =
 \frac{1}{\sqrt{2}}(t \pm x)$, $\pa_{\pm} =
 \frac{1}{\sqrt{2}}(\pa_{t} \pm \pa_{x})$. } The two
 equations are not independent since $\pa_{-}J_{+}=
 -g\pa_{+}J_{-}g^{-1}$.
To derive the Toda theory, we impose the constraints
\footnote{We use a different convention compared to ref.
 \cite{ORphysrep} : $T^{+} = M_{-}$, $T^{-} = -M_{+}$, 
and $T^{0} = -M_{0}$. This convention is chosen in order
 to facilitate comparison with ref. \cite{Brink} later on.  }
\be
	\ga_{\al_{\pm}} = \la E_{\al_{\mp}},J_{\pm} - \ka T^{\pm}
 \ra \approx 0, \label{constr}
\ee
where $E_{\al_{\pm}}$ form an orthogonal set of basis elements of
 ${\cal G}_{\pm}$, and  $\la A,B\ra= \tr(AB)$. The currents are 
thus constrained to have the form
\bea
	J_{\pm}(z)=\ka T^{\pm} + j_{\pm}(z)&,& j_{\pm}(z)\in
 ({\cal G}_{0}+{\cal G}_{\mp})\,. 
\eea
It can be proven that the constraints (\ref{constr}) are
 consistent 
with the dynamics of the WZNW model. The Poisson bracket between 
the components of the currents is given as 
\be
	\{ \la \al,J_{+}(z)\ra ,\la \beta,J_{+}(w)\ra \} 
=\la  [\al,\beta] , J_{+}(z) \ra \de (\zp - w^{+}) +
 \ka\la\al,\beta \ra \pap \de (\zp - w^{+}) \, , \label{pb}
\ee
where $\al$ and $\beta$ are arbitrary elements of ${\cal G}$.
 We will in most of the rest of the text set $\ka=1$, and
 occasionally denote $J_{+}$ by $J$.
By computing the Poisson brackets between the constraints we
 see that the constraints are first class, e.g.
\be
	\{\ga_{\al_{+}},\ga_{\beta_{+}}\} =
 \la [E_{-\al},E_{-\beta}],J\ra + \la E_{-\al},E_{-\beta}\ra
 \approx 0\,,
 \label{frst}
\ee
where, in the first step, we have used (\ref{constr}) and
 (\ref{pb}); we have also suppressed the delta functions.
 In the second step we have used  $\la E_{-\al},E_{-\beta}\ra = 0$,
 and the fact that $[E_{-\al},E_{-\beta}]\in {\cal G}_{m<2}$;
 then since $J \approx T_{+} + j$, where
  $j\in \gn\oplus{\cal G}_{-}$,
 it follows that $ \la [E_{-\al},E_{-\beta}],J\ra\approx 0$,
 by using the properties of the trace (\ref{norm}). 

 We now assume that we can make a generalized Gauss decomposition
 of the form $g=g_{-}g_{0}g_{+}$, where 
$g_{0}=e^{-\phi}$, $\phi = \sum \vphi_{n}h_{n}$, and $g_{\pm}$ 
are the parts of the group which correspond to the parts
 ${\cal G }_{\pm}$ of the algebra. More precisely,
 we assume that this can be done locally (close to the identity),
 this means that we neglect global effects of the group. 
Because of these global effects the constrained WZNW theory
 is larger than the associated Toda theory. For a discussion
 about the global effects, see \cite{Feher}.  Using the
 Gauss decomposition we see that, after a similarity 
transformation
 with $g_{-}$, the equation of motion $\pam J_{+} = 0$, can be 
written as a zero commutation relation of the form
\be
 	\lb \pap - {\cal A}_{+}, \pam - {\cal A}_{-} \rb = 0\,.
\ee
Here ${\cal A}_{-} = -g_{-}^{-1}\pam g_{-}$, and
 ${\cal A}_{+} =\pap g_{0}g_{0}^{-1} +
 g_{0}\pap g_{+}g_{+}^{-1}g_{0}^{-1}$.
We can impose the constraints directly into the equation 
of motion.
 As before, this procedure is consistent since the WZNW
 dynamics do
 not affect the constraint surface. We obtain 
\cite{ORphysrep} ${\cal A}_{+}=-\pap\phi + T^{+}$, and
 ${\cal A}_{-}= e^{-\phi}T^{-}e^{\phi}$. The equation of
 motion then
 becomes
\be
 	\pam\pap\phi = [T^{+},e^{-\phi}T^{-}e^{\phi}]\,,
 \label{eqof}
\ee 
which can be shown to be equivalent to the usual Toda equations of
 motion. We now proceed to show this equivalence. We will make
 use of a few relations and definitions from the theory of
 Lie algebras; we will be brief. For more details, see e.g. 
\cite{hump}.
Instead of the basis used earlier it will now be convenient
 to use a Chevalley basis with the following 
normalizations\footnote{The same notation `$E_{\al}$'
 is used for other basis elements in later sections,
 hopefully no confusion will occur.}
\be \begin{array}{lccl}
	[E_{\al},E_{-\beta}] &=& \de_{\al\beta}H_{\al} & 
 \al\in\Delta_{+} \non \\ 
	{[} H_{\al},E_{\beta}] &=&
 2\frac{(\beta,\al)}{(\al,\al)}E_{\beta} &
 \al,\beta\,\, {\rm roots} \non \\
	{[} H_{\al},E_{\pm\beta}] &=&
 \pm K_{\beta\al}E_{\pm\beta} & \al , 
\beta\in\Delta_{+} \,. \label{bas} \end{array}
\ee
$\Delta_{+}$ denotes the set of (positive)
 simple roots, and $K_{\al\beta}$ is the Cartan
 matrix. The generators of the principal $sl_{2}$
 embedding can, in the basis (\ref{bas}), be written

\be
	T^{0} = \frac{1}{2}\sum_{\al\in
 \Phi_{+}}H_{\al},\,\,\,  T^{+} =
 \sum_{\al\in \Delta_{+}}c_{\al}E_{\al},\,\,\, 
  T^{-} = -\sum_{\al\in \Delta_{+} }E_{-\al}\,,
\ee
where $c_{\al} = 2\sum_{\beta\in\Delta_{+}}K^{-1}_{\al\beta}$,
 and $\Phi_{+}$ is the set of positive roots. The Weyl
 vector, $\rho$, is defined as half the sum of the positive 
roots, i.e.
\be 
	\rho = \frac{1}{2}\sum_{\al\in\Phi_{+}}\al\,.
\ee
$T^{0}$ is the Cartan subalgebra element corresponding 
to $\rho$ under the isomorphism between the Cartan 
subalgebra and the root space. The Weyl vector have
 the following important property $(\rho,\al^{\vee}) = 1$,
 where $\al^{\vee} = 2\frac{\al}{(\al,\al)}$,
 and $\al\in\Delta_{+}$. The dual Weyl vector is 
defined through
\be 
	\rho^{\vee} = 
\frac{1}{2}\sum_{\al\in\Phi_{+}}\al^{\vee} =
 \frac{1}{2}\sum_{\al\in\Phi_{+}}\frac{2\al}{(\al,\al)}\,.
\ee
In general, $\rho \neq \rho^{\vee}$. The dual Weyl vector
 satisfies $(\rho^{\vee},\al)=1$. Using the above
 definitions and properties it can be checked that
 $T_{0}$, and $T_{\pm}$ satisfy the commutation
 relations (\ref{comm}). Expanding $\phi$ in the 
following manner: $\phi = 
\sum_{\al\in\Delta_{+}}\vphi_{\al}H_{\al}$,
 and using the Baker-Hausdorff formula,
 the equation of motion (\ref{eqof}) can be seen to
 reduce to 
\be
	\pap\pam\vphi_{\al} = 
 e^{-\sum_{\beta}K_{\al\beta}\vphi_{\beta}}\,, \label{toda}
\ee
which are the equations of motion of the usual finite
 dimensional Toda theory \cite{Leznov-Saveliev},
 in a suitable normalization. To obtain the form (\ref{toda})
 we also had to make a redefinition of the fields,
 viz. $\vphi_{\al} \rar -\vphi_{\al} -
 \sum_{\beta}K^{-1}_{\al\beta}(\ln(\sum_{\ga}2K^{-1}_{\beta\ga}))$.

We now return to our main discussion.
The equation of motion (\ref{eqof}) can be 
obtained from an effective action

\bea
	S_{\rm eff} = S_{{\rm W}}(g_{0}) + 2\int 
\tr (T^{+}e^{-\phi}T^{-}e^{\phi}) d^{2}z = \non \\
	=\int \tr(\pap\phi\pam\phi)d^{2}z + 
 2\int \tr (T^{+}e^{-\phi}T^{-}e^{\phi}) d^{2}z\,. \label{eff}
\eea
The first class constraints generate infinitesimal
 gauge transformations 
\be \de_{\al} A = \{\int\ga_{\al}(z)dz, A\}\,,
\ee
where $A$ is arbitrary. In particular, the
 currents have the (infinitesimal) gauge
 transformations on the space of classical solutions
\be
	\de J = [\al(\zp),J] + \pap\al(\zp)\,, \label{gt}
\ee
where $\al\in\gm$. Notice that the transformations
 (\ref{gt}) do not affect the Toda field $\phi$,
 which thus is gauge invariant. The {\emt} is given
 by the standard Sugawara expression
\be
	 T_{++}(z) = \frac{\left< J(z)J(z) \right>}{2}. 
\ee
The Sugawara {\emt} is not gauge invariant, this can
 be rectified by adding an improvement term

\be
	\Th_{++}(z) = T_{++}(z) + \left< T^{0}\pap J(z) \right>\,.
\ee

\setcounter{equation}{0}
\section{Lagrangean Realization of the Reduction} \label{sgau}
There is also an equivalent lagrangean realization 
of the WZNW $\rar$ Toda reduction, as a gauged WZNW
 model. We now proceed to describe this approach.
 We use the left-right asymmetric gauge transformations,
\bea
	g \rar hg\bar{h}^{-1} \,,& h=e^{\al(\zp ,\zm)}\,,&
 \bar{h}^{-1}=e^{-\bet(\zp,\zm)}, \label{gaug}
\eea
where $\al\in\gm$, $\bet\in\gp$. The canonical choice for
 a gauge invariant action is
\be
	S(A_{-},A_{+},g) = S_{{\rm W}}(g) +
 2 \int\tr[ A_{-}(\pap gg^{-1}) + A_{+}(g^{-1}\pam g)
 + A_{-}gA_{+}g^{-1}] d^{2}z\,. \label{gw}
\ee
This action is invariant under the gauge
 transformations (\ref{gaug}), together with

\bea
	A_{-} \rar e^{\al}A_{-}e^{-\al} +
 e^{\al}(\pam e^{-\al}) &,& A_{+}
 \rar e^{\bet}A_{+}e^{-\bet} +  (\pap e^{\bet}) e^{-\bet}\,. 
\label{gau}
\eea

The action (\ref{gw}) is ``derived'' in the usual way;
 the natural starting point is $S_{{\rm W}}(hg\bar{h}^{-1})$,
 which may be rewritten using the Polyakov-Wiegmann 
identity \cite{Polyakov-Wiegmann83} \cite{Polyakov-Wiegmann84}

\bea
	S_{{\rm W}}(abc)&=&S_{{\rm W}}(a)+S_{{\rm W}}(b)+
S_{{\rm W}}(c) +
 2\int\tr(a^{-1}\pam a)(\pap bb^{-1})d^{2}z +\\ && +
 2\int\tr[ (b^{-1}\pam b)(\pap cc^{-1})+
(a^{-1}\pam a)b(\pap c)c^{-1}b^{-1}] d^{2}z\,.
\eea 
(Notice that $S(h)=0$, and $S(\bar{h}^{-1})=0$, because
 of the properties of $\tr$). By making the identification
 of the gauge fields according to, $A_{+} =
 \pap \bar{h}^{-1}\bar{h}$, $A_{-} = h^{-1}\pam h$,
 the gauge invariant action (\ref{gw}) follows. 
We have potential problems with anomalies; an
 infinitesimal gauge variation of $S$ gives
\be
	\de S = 2\tr(\beta \pam A_{+} - \al \pap A_{-})\,.
\ee
The variation vanishes, however, as a consequence
 of the properties of the trace. The above action 
is not quite what we want. We want to constrain 
some of the components of the WZNW currents to be
 constant (not zero). To achieve this feature we
 add a gauge invariant term to $S$, viz.
\be 
	-2\int\tr(A_{-}T_{+} - A_{+}T_{-})d^{2}z\,.
\ee
It can be checked that this transforms as a total
 derivative under an infinitesimal gauge transformation.
 Summarizing, we get

\be
	S = S_{{\rm W}}(g) + 2\int\tr[ A_{-}(\pap gg^{-1}
 - T_{+}) + A_{+}(g^{-1}\pam g + T_{-}) +
 A_{-}gA_{+}g^{-1}] d^{2}z\,,
\ee
which is invariant under the  gauge transformations 
(\ref{gau}).
We can use the gauge freedom to set $A_{\pm}=0$.
If we eliminate the $A$'s using their Euler-Lagrange
 equations and remember the gauge choice, we
 recover the constraints (\ref{constr}), and the 
equations of motion of the WZNW model. This does 
not, however, completely fix the gauge, we still 
have the freedom to make gauge transformations of
 the form $g \rar e^{\al(\zp)}ge^{-\bet(\zm)}$. 
This will give us back the gauge transformations
 (\ref{gt}).
\setcounter{equation}{0}
\section{The General Algebra} \label{alg}
In section (\ref{wzw}) we based our discussion on
 a simple maximally non-compact Lie algebra.
In this section we will describe a generalization 
of the simple Lie algebras $sl_{N}$ and the associated
 generalized Toda theories. The algebra we have in
 mind originated in higher spin theories \cite{Vasiliev89}
 \cite{Vasiliev91}, and is defined as the Lie algebra
 obtained from the (associative) algebra of all 
monomials of two operators, $a^{\pm}$, satisfying 
the commutation relations
\bea
	[ a^{-},a^{+} ] &=& 1 + 2\nu K \,, \non \\
 \{ K,a^{\pm}\} &=& 0 \;\; ,\;\;\;\;  K^{2}=1\,.
\eea
Here $\nu$ is a free parameter. A general element 
of the algebra can be written as
\be
B = \sum_{A=0}^1 \sum_{n=0}^\infty \frac{1}{n!}
b^A_{\, \alpha_1 \ldots \alpha_n } K^A\,
a^{\alpha_1} \ldots a^{\alpha_n }\,.\label{gel}
\ee
We can choose $b^A_{\, \alpha_1 \ldots \alpha_n }$ 
to be totally symmetric in the lower indices, so the 
basis elements can be chosen to be the Weyl ordered 
products
\be
\label{base}
E_{nm}=\Big ( (a^+ )^n (a^- )^m \Big )_{\rm Weyl} =
\frac{1}{(n+m)!}
\Bigl ( (a^+ )^n (a^- )^m +((n+m)!-1)  )\,
\mbox{permutations}\Bigr ) \,,
\ee
together with the elements $KE_{nm}$. We would like 
to point out that in general there does not exist a 
generalization of the Chevalley basis described in 
section \ref{wzw} to the infinite-dimensional algebra
 considered in this section.
The quadratic combinations of $a^{\pm}$ 
\bea
	T^{\pm} = \frac{1}{2}(a^{\pm})^{2} &,& T^{0}
 = \frac{1}{4}\{a^{+},a^{-}\} \,, 
\eea
satisfy the commutation relations (\ref{comm}), and 
thus form an $sl_{2}$ subalgebra. 
This realization of $sl_{2}$ dates back to Wigner
 \cite{Wigner50}, see also \cite{Deser} \cite{Sudarshan}.
 The algebra under discussion can be shown to be 
isomorphic to the universal enveloping algebra of
 $osp(1,2)$, $U(osp(1,2))$, divided by a certain 
ideal \cite{Bergshoeff91}.  When $d=N$, where 
$d=\frac{2\nu +1}{2}$ and $N$ is a positive integer, 
the algebra becomes, after dividing out an ideal, 
isomorphic to $gl(N,N-1)$, when considered as a Lie algebra.
The algebra has a unique trace operation 
\cite{Vasiliev89} \cite{Vasiliev91}, defined as
\be
\label{str}
str (B) = b^0 -2\nu b^1 \,,
\ee
with the properties,
\be
\label{strprop}
str (AB) =
(-1)^{\pi(A)}str (BA)=(-1)^{\pi(B)}str(BA) \,,
\ee
where $\pi(A)$
is defined to be equal to $0$ or $1$ for monomials with
 even or odd powers $n$ in (\ref{gel}), respectively. In
 the bosonic subalgebra (spanned by all even monomials, 
i.e. all $E_{nm}$'s, with $n+m={\rm even}$) we can use 
the projection operators 
${\rm P\!}_{\pm} = \frac{1 \pm K}{2}$ to split the algebra 
into two parts. The two sectors carry the same amount of 
information so without loss of generality we will in the
 sequel only work in the $\pr$ sector of the bosonic subalgebra.
 The element $T^{0}$ defines an integral gradation of the
 bosonic algebra, hence the algebra can be split as (${\cal G}$
 denotes the bosonic algebra)
\be
	{\cal G} = \gm\oplus\gn\oplus\gp\,,
\ee
where the $\gp$ ($\gm$) are the parts of the algebra with
 positive (negative) grade with respect to $T^{0}$, and
 $\gn$ is the zero grade part. {}From now on we write $\tr$
 instead of $str\pr$. It can be proven \cite{Vasiliev91} that 
\be
	\tr(E_{nm}E_{rs}) = \de_{mr}\de_{ns}f_{nm}(\nu),
\ee
where $f_{nm}(\nu)$ is a certain function of $\nu$.
 The bilinear form $\la A,B \ra = \tr(AB)$ is not positive
 definite in general. When $d = N$, where $N$ is a
 positive integer, the bilinear form $\la\cdot,\cdot\ra$
  becomes degenerate; all $E_{nm}$'s with $n+m\geq 2(N+1)$ 
have $f_{nm} = 0$, and
 hence decouple under the trace; they form the ideal 
mentioned earlier. The ideal can be divided out, leaving
 us with a finite set of basis elements for which the
 bilinear form is non-degenerate. In fact, the residual
 set of elements form the algebra $gl_{N}$.
In order to make the correspondence with the formulas
 is section \ref{wzw} as close as possible we could
 use the following notation for the basis elements: 
$E^{s}_{n}= E_{kl}$, with $n = \frac{k+l-|k-l|}{2}$,
 and $s = \frac{k-l}{2}$.
In the following sections we will make use of the operator $t$,
 defined through
\bea
	 [T^{+},t(x)] &=& x - \Pi_{-}(x) \,, \non \\
	 t([T^{+},x]) &=& x - \Pi_{+}(x)	\,. \label{tinv}
\eea
Here $\Pi_{-}$ ($\Pi_{+}$) is the projector onto the subset
 consisting of lowest (highest) weight elements, and $x$ is 
an arbitrary element in the algebra. $t$ lowers
 the grade one step, and when $t$ hits a lowest weight
 element it gives zero. We will also use the following relation
\be
	 \la At(B)\ra = -\la t(A)B\ra\,. \label{tprop}
\ee
The action of $t$ on the basis elements is
\bea
	t(E_{n,m}) = -\frac{1}{m+1}E_{n-1,m+1}&,&t(E_{0,m}) = 0\,.
\eea
We will in what follows assume that the ideal $\pr E_{0}^{0}$
 has been factored out, which implies that we get $sl_{N}$
 when $d=N$. We will now make contact with section \ref{wzw}.
 Let us first note that in section \ref{wzw} (most of) the
 results did not depend on the specific form of the Lie
 algebra used. It only depended on the fact that the Lie
 algebra had an $sl_{2}$ subalgebra, which induced an 
integral gradation of the algebra. We can thus use the
 $\pr$ part of the bosonic sector of the algebra 
described above as a basis for constructing a generalized
 Toda theory.
We have one problem though: As always, it is
 problematic to integrate an infinite-dimensional
 algebra (representation) into a group. In
 the infinite-dimensional case  we do not in 
general have the simple relation (exponentiation)
 between the algebra and a group, so whether 
we have an 'WZNW' action which gives the equations
 of motion in the general case is an interesting problem.
 We will not address this problem here, we will 
simply base our discussion on the equation of motion,
 together with the (first class) constraints
 (\ref{constr}) which utilises only the algebra.
 Using this approach we obtain the model of ref. 
\cite{Brink} by noting that the (effective) action
 (\ref{eff}) and the associated equation of motion 
(\ref{eqof}) coincide with the ones used in \cite{Brink}, 
if the underlying algebra is taken to be the (bosonic) 
algebra described in this section. 

\setcounter{equation}{0}
\section{Different Gauges in the Reduced WZNW model} \label{diff}

We now return to the discussion of the WZNW $\rar$ Toda
 reduction, with the understanding that the underlying 
 Lie algebra is the $\pr$ part of the bosonic sector of
 the algebra described in the previous section (with the
 ideal $\pr E^{0}_{0}$ factored out), which encompasses
 e.g. $sl_{N}$ as a special case. 
As discussed in sections \ref{wzw} and \ref{sgau} we 
have a gauge invariance in our system (cf. (\ref{gt})).
  The generators of the ${\cal W}$ symmetry are gauge 
invariant polynomials in the components of the WZNW
 current $J$; they form a closed algebra.  There are 
several possible ways of fixing the gauge. A useful 
gauge choice is the lowest\footnote{This is the
 highest weight gauge of \cite{ORphysrep}, in the
 conventions we use.} weight gauge. 
The lowest weight gauge is defined by gauge fixing
 $j = J - T^{+}$ to lie in $\ker ad_{T^{-}}$. This 
completely fixes the gauge. The lowest weight gauge
 has a remarkable feature  \cite{ORphysrep} which 
makes the ${\cal W}$ algebra particularly transparent,
 viz. if we write 
\be
	J = T^{+} + \sum_{i}
 j_{i}\frac{(T^{-})^{i}}{\la (T^{+})^{i}(T^{-})^{i}\ra}
\ee
 then the $j_{i}$'s form a basis of primary fields of 
the ${\cal W}$ algebra (with the exception of $j_{1}$,
 which is the \emt; here we have the usual central
 charge). That is, on the constraint surface, we have 
$W_{i}=j_{i}$, where $W_{i}$ is the $i$th generator 
of the ${\cal W}$ symmetry.

 Another gauge choice of interest is the diagonal
 gauge. This gauge is specified by demanding that
 $j$ should lie in $\ker ad_{T^{0}}$. Thus, in this
 gauge we can write $J(z) = T^{+} - \pap\ta$,
 where $\ta\in\gn$. Notice that $\ta$ here is {\em not}
 the Toda field, but a free field. In the diagonal
 gauge the Dirac bracket is equal to the original
 Poisson bracket, as was shown in ref. \cite{ORphysrep},
 using methods similar to those we use in section 
\ref{sgov}. Summarizing, the field $\ta$ satisfies canonical
 commutation relations, as well as $\pam\pap\ta = 0$.
 The ${\cal W}$ generators are traces of 
polynomials in the components of the current $J$
, i.e. polynomials in $\vth_{n}$,
 if we use the expansion $\ta = \sum_{n}\vth_{n}h_{n}$.

 An important issue is whether the gauges are
 accessible.
 The lowest weight gauge is accessible, as was 
shown in ref. \cite{ORphysrep}; the diagonal gauge
 on the other hand is only locally accessible,
 this however is sufficient for our purposes 
since we will only be concerned with calculating
 Dirac brackets which are local objects. 

We will now derive an (explicit) expression for
 the Dirac bracket in the lowest weight gauge. 
For convenience we will, in this and the following
 section, use $\frac{E_{nm}}{\sqrt{f_{nm}(\nu)}}$
 as our basis elements, with the understanding that
 when $d=N$ only the finite number of elements with
 $f_{nm}\neq 0$ should be included. The rest automatically
 decouples under the trace operation. The fact that the
 rescaling makes some of the basis elements imaginary is
 not a problem considering the way that they enter into 
the definition of the constraints. The end result will 
be insensitive to the rescaling of the basis elements. 
We will in what follows denote the rescaled basis elements
 by $E_{\al}$. They satisfy
\be
	 \tr(E_{\al}E_{-\beta}) = \de_{\al\beta}\,.
\ee
Here $E_{-\beta}$ is the dual of $E_{\beta}$, which
 means that if $E_{\beta}$ equals 
$\frac{E_{nm}}{\sqrt{f_{nm}(\nu)}}$, then
 $E_{-\beta}$ equals $\frac{E_{mn}}{\sqrt{f_{nm}(\nu)}}$.
 Recall that the first class constraints were given by
\be
	 \ga_{\al} = \la E_{\al},J - T^{+} \ra \approx 0\,,
\ee
for all $E_{\al} \in\gm$. Together with the gauge fixing
 conditions $c_{\beta} = \la E_{\beta},J-T^{+}\ra$,
where $E_{\beta}\in\gn\oplus\gp$ but is not highest
 weight, the constraints become second class. These 
constraints can be written
\be
	 \chi_{\al}=\la tE_{\al},J-T^{+}\ra \approx 0\,,
 \label{sec}
\ee
for all (except lowest weight) basis elements $E_{\al}$ 
of the algebra. The operator $t$ has been defined in
 section \ref{alg}. On the constraint surface
 $\chi_{\al} = 0$, $J$ will thus be of the
 form $J = T^{+} - \mu$, where 
$\mu \in \ker ad_{T^{-}}$. We now recall that
 the Dirac bracket is given by 
\be
	\{ \cdot,\cdot \}^{*} =
 \{ \cdot,\cdot\} -
 \sum_{\al\beta}\int\!\!\!\int\{\cdot,\chi_{\al}(x)\}
C_{\al\beta}^{-1}(x,y)\{\chi_{\beta}(y),\cdot\} dxdy\,,
\ee
where $C_{\al\beta}(x,y) = \{\chi_{\al}(x),\chi_{\beta}(y)\}$.
 Using the explicit expression for the constraints (\ref{sec})
 and the commutation relations (\ref{pb}), together with 
(\ref{tprop}) and (\ref{tinv}) we get, on the constraint
 surface $\chi_{\al}=0$
\be
	 C_{\al\beta}(x,y) = \la E_{\al}
tU_{0}^{-1}E_{\beta} \ra\de(x-y)\,.
\ee
Here $E_{\al}$ and $E_{\beta}$ are basis elements
 of the algebra which are not lowest weight, and
  $U_{0}^{-1} = 1 - D_{0}t$, where
 $D_{0} = \pap + [\mu,\cdot]$. We need
 to calculate the inverse of the matrix, 
$\la E_{\al}tU_{0}^{-1}E_{\beta} \ra$. We 
notice that $t$ does not have an inverse, 
the closest we can get is (\ref{tinv}). 
 However, when sandwiched between basis 
elements as below, we can effectively use 
$t^{-1} = [T^{+},\cdot]$. We will only use
 $t^{-1}$ as a short-hand notation for $[T^{+},\cdot]$.
 We now show that the inverse of 
$\la E_{\al}tU_{0}^{-1}E_{\beta}\ra$ is given by
 $\la E_{-\beta}U_{0}t^{-1}E_{-\al} \ra$, where 
$E_{-\al}$ and $E_{-\beta}$ are not highest weight
 elements, explicitly
\be
\sum_{\beta}\la E_{\al}tU_{0}^{-1}E_{\beta} \ra\la
 E_{-\beta}U_{0}t^{-1}E_{-\ga} \ra = \de_{\al\ga}\,, 
\label{cc}
\ee
where $E_{\al}$ is not lowest weight and $E_{-\ga}$
 is not highest weight. The sum in (\ref{cc}) runs
 over values such that the basis elements $E_{\beta}$
 are not lowest weight, but the sum can be extended
 to the entire algebra, as follows from the fact
 that the terms where $E_{\beta}$ is lowest
 weight automatically are zero, because of the 
$t$ in $tU_{0}^{-1}$. We can then use the closure
 relation (the sum runs over all $\beta$)

\be
 	\sum_{\beta} E_{\beta}\ra\la E_{-\beta} = 1\,.
\ee
Equation (\ref{cc}) then follows, since (\ref{tinv})
 implies that $\la E_{\al}tt^{-1}E_{-\ga}\ra = \de_{\al\ga}$,
 when $E_{\al}$ is not lowest weight and $E_{-\ga}$ is not 
highest weight.
$C^{-1}_{\al\beta}(x,y)$ is defined through
\be
	 \sum_{\ga}\int C_{\al\ga}(x,z)
C^{-1}_{\ga\beta}(z,y)dz = \de_{\al\beta}\de(x-y)\,.
\ee
Hence, using the above results we get
\be
	 C_{\al\beta}^{-1}(x,y) =
 \la E_{-\al}U_{0}t^{-1}E_{-\beta}\ra\de(x-y)\,.
\ee
We are now in a position to calculate the Dirac bracket
 between the components of the $\mu$-field. Since the
 components of the $\mu$-field are the generators of 
the ${\cal W}$ algebra we will thus obtain an expression
 for the commutators of the ${\cal W}$ algebra. We 
would like to stress that although we derive all formulas
 in the general case they apply also to the finite-dimensional
 cases (ordinary $A_{N}$ Toda theories and ${\cal W}_{N}$
 algebras); we only have to choose the parameter $\nu$ 
appropriately. The basis elements can in these cases
 be realized as finite dimensional matrices. To
 calculate the Dirac brackets we start by recalling
 that $\mu = -J + T^{+}$, which gives 
\bea
	\{\int\la u,\mu\ra dz,\chi_{\al}\} &=&
 -\la tD_{0}u,E_{\al}\ra\,, \non \\       
 \{\chi_{\beta},\int\la v,\mu\ra dw\} &=&
 \la E_{\beta},tD_{0}v\ra\,, \label{uv}
\eea
where we have used (\ref{sec}), (\ref{tprop})
 and (\ref{tinv}) together with the ubiquitous
 (\ref{pb}). In (\ref{uv}), $u$ and $v$ are
 arbitrary highest weight elements, and
 $E_{\al}$, $E_{\beta}$ are not lowest weight
 elements. The final piece of information we 
need is, $\{\la u,\mu\ra ,\la v,\mu\ra\} = 0$.
 Collecting the above results together, we get
 the following expression for the Dirac
 brackets between the components of the $\mu$-field

\bea
	\{ \int\la u,\mu\ra dz ,
\int\la v,\mu\ra dw \}^{*}& =&
 \int\sum_{\al\beta} \la tD_{0}u,E_{\al}\ra\la E_{-\al}U_{0}
t^{-1}E_{-\beta}\ra\la E_{\beta}tD_{0}v\ra = \non \\ &=& 
\int\la uD_{0}tU_{0}D_{0}v\ra  = \int\la uU_{0}D_{0}v\ra. 
\label{lwdb}
\eea
We now outline the method used to prove the steps in
 (\ref{lwdb}). The sums run over all $\al$ and $\beta$
 such that $E_{\al}$,$E_{\beta}$ never are lowest
 weight elements.  To show the second equality we
 insert $U_{0}^{-1}\sum_{\ga}E_{\ga}\ra\la 
E_{-\ga}U_{0} = 1$ between $t$ and $D_{0}$ in
 the third bracket, and use
\be
	\sum_{\beta}\la E_{-\al}U_{0}t^{-1}E_{-\beta}
 \ra\la E_{\beta}tU_{0}^{-1}E_{\gamma} \ra = 
\de_{\al\ga}\,, \label{ag}
\ee
which is true if $E_{\ga}$ is not lowest weight,
 otherwise the result is zero. The sum in 
(\ref{ag}) runs over all $\beta$ such that $E_{\beta}$
 is not a lowest weight element, but it is easy to see
 that it can be extended to the entire algebra. The 
sum which remains after (\ref{ag}) has been used can 
similarly also be extended to the whole algebra.
The final equality in (\ref{lwdb}) follows from 
$\la u,D_{0}v\ra = 0$, together with the definition
 $U_{0} = (1-D_{0}t)^{-1}$, which implies that 
$D_{0}tU_{0} = U_{0} -1$.
The result we have obtained is in  agreement with
 the expression for the $\mu$-bracket derived in 
\cite{Brink}. Another way to obtain this bracket 
is through use of the governing equation of the 
next section; this path was the one followed in 
ref. \cite{Brink}. The explicit expression for 
the Dirac bracket furnishes us with an algorithmic
 method for computing the ${\cal W}$ algebra.
The ${\cal W}$ algebra under discussion was
 denoted ${\cal W}^{\cal G}$ in \cite{ORphysrep}.
 In our case ${\cal G}$ is the algebra of section \ref{alg}.

\setcounter{equation}{0}
\section{The Governing Equation} \label{sgov}

In this section we will be concerned with the
 connection between the diagonal gauge, and the 
lowest weight gauge. In the diagonal gauge the Dirac
 bracket is simple (equal to the original Poisson
 bracket), but the ${\cal W}$ generators are complicated 
(polynomials in the components of the current).
 In the lowest weight gauge, on the other hand,
 we have the opposite situation, here the ${\cal W}$ 
generators are simple (equal to the components the 
current, see the previous section), but the Dirac
 bracket is nonlinear.

In order to establish a connection between the
 two gauges we will exploit the fact that the 
${\cal W}$ currents are gauge invariant. The 
first class constraints are as before given by
\be
	 \ga_{\al} = \la E_{\al},J - T^{+} \ra 
\approx 0\,,
\ee
for all $E_{\al} \in\gm$.
We make a partial gauge fixing of the first class
 constraints. We gauge-fix all constraints $\ga_{\al}$,
 which are not of the form $\la th,J-T^{+}\ra$,
 where $h$ is an arbitrary zero grade element. 
The gauge fixing conditions are chosen to be, 
$c_{\al} = \la tE_{\al},J-T^{+}\ra$ where
 $tE_{\al}\notin\gn$, and $E_{\al}$ is not a 
lowest weight basis element. The second class 
constraints are thus
\be
 	\chi_{\al} = \la tE_{\al},J-T^{+}\ra \approx 0\,, 
\label{chi}
\ee
where $tE_{\al}$ is not of the form $h$ or $th$ (where
 $h$ is an arbitrary grade zero element) and $E_{\al}$
 is not a lowest weight basis element. We notice that 
we can reach both the diagonal and the lowest weight
 gauge by further gauge fixing the remaining first 
class constraints\footnote{The fact that these really
 are first class constraints can be shown by computing
 their Dirac Brackets, $\{\vrho_{\al},\vrho_{\beta}\}^{*}
\approx 0$.}
\be
	 \vrho_{\al} = \la th_{\al},J-T^{+}\ra \approx
 0\,, \label{rem}
\ee
where $h_{\al}$ form a set of orthonormal basis elements
 of $\gn$. On the constraint surface we can write
 $J = T^{+} -\pap\ta - \mu$. To be precise this is
 a slight abuse of notation. What we have called
 $\mu$ here really only corresponds to $\mu$ in
 the previous section if we set $\ta=0$ i.e. if we
 go to the lowest weight gauge. A similar statement
 holds for $\ta$. In analogy with the calculation
 in the previous section we can derive the following
 expression, on the constraint surface, for the
 constraint matrix (suppressing the delta function)
\be
	 C_{\al\beta} =  \{\chi_{\al},\chi_{\beta}\} =
 \la E_{\al}tU^{-1}E_{\beta}\ra\,.
\ee
The range of indices in $C_{\al\beta}$ are such that 
$E_{\al}$, $E_{\beta}$ do not have grade $0$ or $1$,
 and are not lowest weight elements. We have introduced 
the notation $U^{-1} = 1 - Dt$, where 
$D = \pap + [\pap\ta,\cdot] + [\mu,\cdot]$.  
The constraint matrix has the generic block form,

\be
	C_{\al\beta} = \left( \begin{array}{cc} 0 & A
 \\  -A^{T} & B \end{array} \right).
\ee
The matrix elements are labeled by $\al$ and $\beta$ 
in $C_{\al\beta}=\la E_{\al}tU^{-1}E_{\beta}\ra$.
 When we say e.g. that $\al$ has zero grade we really
 mean that $E_{\al}$ has zero grade. This rule of using
 $\al$ and $E_{\al}$ interchangeably will occasionally
 be used in the remaining part of this section. The range
 of the indices in
 $A_{\al\beta}=\la E_{\al}tU^{-1}E_{\beta}\ra$, are such
 that $\al$ has grade $\leq -1$ (but is not lowest weight),
 and $\beta$ has grade $\geq 2$.  The inverse of the 
constraint matrix is

\be
	C^{-1}_{\al\beta} = \left(
 \begin{array}{cc} (A^{T})^{-1}BA^{-1} &
 -(A^{T})^{-1} \\ A^{-1} & 0 \end{array} \right).
 \label{cinv}
\ee
Thus, in order to be able to calculate the inverse
 we need  $A^{-1}$. The inverse of $A_{\al\beta}$ 
is given by, $A^{-1}_{\al\beta} = 
\la E_{-\al}Ut^{-1}E_{-\beta}\ra $, where  $E_{-\al}$
 has grade $\leq -2$, and $E_{-\beta}$ has grade
 $\geq 1$ but is not highest weight. To show that 
\be
	\sum_{\beta}A_{\al\beta}A^{-1}_{\beta\ga}
 = \sum_{\beta}\la E_{\al}tU^{-1}E_{\beta}\ra\la 
E_{-\beta}Ut^{-1}E_{-\ga}\ra =  \de_{\al\ga}\,, \label{aa}
\ee
where the sum runs over all $\beta$ with grade
 $\geq 2$, we use the fact that $tU^{-1}$ decreases
 the grade when acting on elements which are not 
lowest weight. Because of this property $A_{\al\beta}$ 
is zero unless the grade of $\beta$ is $\geq 2$
 (since $\al$ has grade $\leq -1$), using the 
properties of the trace. The sum over $\beta$ 
in (\ref{aa}) can thus be extended to the entire
 algebra. Finally, the closure relation can be 
invoked, and equation (\ref{aa}) follows.

The remaining first class constraints generate 
gauge transformations. Since the ${\cal W}$ 
currents are gauge invariant, they satisfy 
$\de W = \{\vrho,W\}^{*} \approx 0$, for all 
first class constraints $\vrho$ of the form 
(\ref{rem}). Here $W$ is an arbitrary ${\cal W}$
 current. $W$ is given as a trace of a polynomial
 in $J$, which implies (weakly) that it is a 
function of $\pap\ta$ and $\mu$. This gives us
 (summation over $i$ is understood)

\be
	\de W = \left[ \int\{\vrho,\mu_{i}\}^{*}
 \frac{\de}{\de \mu_{i}} +\int\{\vrho,\pap\vth_{i}\}^{*} 
\frac{\de}{\de \pap\vth_{i}}\right] W\approx 0\,. 
\label{wprop}
\ee 
Here $\mu_{i} = \la E_{i},\mu\ra$ and $\pap\vth_{i} =
 \la h_{i},\pap\vth\ra$, where the set of 
 orthonormal basis elements $E_{i}$ span the subspace of highest
 weight elements, and the orthonormal set $h_{i}$ span $\gn$.
 We have also used the property,
 $\frac{\de\mu_{i}(x)}{\de\mu_{j}(y)} = \de_{ij}\de(x-y)$,
 and analogously for $\pap\vth_{i}$.
 Remembering that $J \approx T^{+} - \pap\ta - \mu$,
 we see that in order to discover the
 implications of the gauge invariance of $W$,
 we need to calculate $\{\vrho,\la h,J\ra\}^{*}$
 and $\{\vrho,\la u,J\ra\}^{*}$, where $h$ has
 zero grade and $u$ is a highest weight element
. Here we choose $\vrho = \la \int t\xi(z) dz,J-T^{+}\ra$ 
where $\xi$ is an arbitrary zero grade element.
 We start with $\{\vrho,\la h,J\ra\}^{*}$. The 
second term in the Dirac bracket is schematically

\be
	\sum_{\al\beta} \{\vrho,\chi_{\al}\}
C_{\al\beta}^{-1}\{\chi_{\beta},\la h,J\ra\}. \label{secdb}
\ee
We now show that (\ref{secdb}) is zero. First
 recall that the second class constraints were 
given by (\ref{chi}). The first bracket in 
(\ref{secdb}) can be calculated to give the
 result $-\la\xi tDtE_{\al}\ra$. Using the fact
 that $t$ lowers the grade, we see that $\al$ is
 constrained to have grade $\geq 2$ (remember that
 $\xi$ is a zero grade element).  Similarly, the 
other bracket can be calculated to give $-\la E_{\beta}tDh\ra$.
 Its implications are to constrain $\beta$
 to have grade $\geq2$ (remember that there are
 no constraints $\chi_{\beta}$, with $E_{\beta}$
 a grade 1 element). Using the block form of 
$C_{\al\beta}^{-1}$ (\ref{cinv}), the result 
follows ($C^{-1}_{\al\beta}$ is zero in the 
appropriate sector).
The first term in the expression for the Dirac
 bracket can be calculated to give
 $\{\vrho,\la h,J\ra\} = \la \xi,h\ra$; hence
\be 
	\{\vrho,\la h,J\ra\}^{*} = \la\xi,h\ra\,.
\ee
We now turn to $\{\vrho,\la u,J\ra\}^{*}$. The
 second term in the Dirac bracket is

\be
	 \int\!\!\!\int dxdy\sum_{\al\beta}
 \{\vrho,\chi_{\al}(x)\}
C_{\al\beta}^{-1}(x,y)\{\chi_{\beta}(y),\la u,J\ra\}.
 \label{secnd}
\ee

Here, as above, the first bracket is non-zero only if
 $\al$ has grade $\geq 2$, so (\ref{secnd}) becomes using
 (\ref{cinv}), and reinstating the delta function

\be
	\sum_{\al\beta} \la\xi tDtE_{\al}\ra\la
 E_{-\al}Ut^{-1}E_{-\beta}\ra\la E_{\beta}tDu\ra.
\ee
Here $E_{-\beta}$ has grade $\geq 1$ (but is not
 highest weight), and $E_{-\al}$ has grade $\leq -2$.
 We can use the same trick as in the previous section,
 and insert $U^{-1}\sum_{\ga}E_{\ga}\ra\la E_{-\ga}U = 1$,
 between $t$ and $D$. The fact that $E_{\beta}$ has grade
 $\leq -1$, means that we can restrict the sum over $\ga$
 to run over values such that $E_{\ga}$ has grade $\geq 2$
 (as  follows from the fact that $tU^{-1}$ decreases the
 grade). We get
\be
	\sum_{\al\beta\ga}
 \la\xi tDtE_{\al}\ra A^{-1}_{\al\beta}
A_{\beta\ga}\la E_{-\ga}UDu\ra =
 \sum_{\al}\la\xi tDtE_{\al}\ra\la E_{-\al}UDu\ra\,.
\ee
Since the first bracket is zero for elements
 $E_{\al}$, which do not have grade $\geq 2$, we see that the sum
 over $\al$ can be extended to the entire algebra. Then, using the
 closure relation and the definition of $U$, we get
\bea
	 \sum_{\al}\la\xi tDtE_{\al}\ra\la
 E_{-\al}UDu\ra\ &=& \la\xi,\sum_{n=2}^{\infty} (tD)^{n}u\ra\,.
\eea
The first term in the expression for the Dirac bracket 
is $\{\vrho,\la u,J\ra\} = -\la \xi,tDu\ra$, including
 this term we get
\bea
	\{\vrho,\la u,J\ra\}^{*} &=& -\la\xi,tDu\ra - 
\la\xi,\sum_{n=2}^{\infty}(tD)^{n}u\ra =\\
 &=&-\la\sum_{n=1}^{\infty}(Dt)^{n}\xi,u\ra  = -\la U\xi,u\ra.
\eea
In the last step we have used $\la \xi,u\ra=0$. We
 have also done an integration by parts, which is 
permissible since our expression is to be integrated 
over, cf. (\ref{wprop}). Collecting, the gauge 
invariance property of the ${\cal W}$ currents 
(\ref{wprop}) can be seen to reduce to the condition

\be
	\left[ \int\la \xi,\frac{\de}{\de\pap\ta}\ra 
- \int\la U\xi,\frac{\de}{\de\mu}\ra \right]W = 0. \label{gov}
\ee
Here we have used the definition $\frac{\de}{\de\mu} =
 \sum_{i}E_{i}\frac{\de}{\de\mu_{i}}$, where the orthonormal set 
 of $E_{i}$'s span the highest weight elements,
 which implies
\be
	\int\la \eta,\frac{\de}{\de\mu}\ra\mu =
 \Pi_{-}(\eta) \,,
\ee
and similarly for $\frac{\de}{\de\pap\vth}$.  We
 see that (\ref{gov}) is the governing equation of
 \cite{Brink} with one important difference. In
 (\ref{gov}) the $\ta$-field is not the Toda field but
 a free field. The governing equation gives the
 connection between the ${\cal W}$ generators in 
the two gauges. Equation (\ref{gov}) was solved in
 \cite{Brink} with the result (P denotes path 
ordering)\footnote{In \cite{Brink} the notation
 $J^{s}$ was used on the left hand side instead 
of $W$, and $j$ was used on the right hand side.},
\be
	W(\pap\ta,\mu) = {\rm P}\exp(\int_{0}^{1}
\left[ \int \la U(s\pap\ta)\pap\ta,\frac{\de}{\de\mu} 
\ra\right] ds)W(\mu).
\ee
Hence, $W(\pap\ta)=W(\pap\ta,\mu)|_{\mu=0}$.
In our case this is a (Miura) transformation,
 which expresses the $W$ generators in terms 
of free fields satisfying canonical commutation
 relations. We once again would like to point 
out that (by choosing the parameter $\nu$ 
appropriately) all results apply also to the
 finite-dimensional cases. How is $W(\pap\ta)$
 related to the ${\cal W}$ currents expressed in
 terms of the Toda field $\phi$, $W(\pap\phi)$? 
{}From the gauge invariance of $W$, it follows that 
$W(\phi)$ and $W(\ta)$ are equal. Notice that $\phi$
 is gauge invariant so $W(\phi)$ is the same in all
 gauges. To further investigate the relation between
 the two realizations of $W$, we first note that
 $J = \pap g_{-}g_{-}^{-1} + g_{-}(T^{+} -
 \pap\phi)g_{-}^{-1}$. We then write $g_{-}=1+\beta$,
 where the infinitesimal element $\beta$ lies in
 $\gm$, and recall that the (infinitesimal) gauge 
transformation of the current is 
\be
	\de J = [\al,J] + \pap\al\,.
\ee
This transformation can then be seen to imply
 $\beta \rar \beta + \al$.  Because the ${\cal W}$
 currents are gauge invariant, we may conclude that
 they do not depend on $\beta$. Since the $z$
 dependent part of $\beta$ can be (almost) 
arbitrarily changed, the fact that $\beta$ 
disappears from $W(\phi)$ must be a consequence 
of the properties of the trace and of
 cancellation between the different terms in $W(\phi)$,
 and not a consequence of the space-time
 dependence of $\beta$.  We can thus set $g_{-}=1$
 when calculating the currents. Notice that
 $g_{-}=1$ is not a possible gauge choice, since then
 $J_{-}$ would not be consistent with the constraints and 
the WZNW dynamics.  We have shown that we have the same
 functional dependence in the two cases i.e. $W[\phi] = W[\ta]$,
 a stronger result than mere equality.
What is the explicit relation between the $\ta$ and
 the $\phi$ fields? We will address this question in
 the simplest possible case, the Liouville model,
 which is the $A_{1}$ Toda theory. The $sl_{2}$ algebra
 is obtained from the general algebra by setting $d=2$.
 The Gauss decomposition can in this case be written
\be
	g = e^{\eta T^{-}}e^{-\vphi T^{0}}e^{\xi T^{+}}\,.
\ee
Here $T^{0}$, and $T^{\pm}$ can be thought of as
 2-dimensional matrices. The currents become
\bea
	 J_{+} &=& \pap\eta T^{-} + e^{\eta T^{-}}(T^{+}
 - \pap\phi)e^{-\eta T^{-}}\non \\
	 J_{-} &=& -\pam\xi T^{+} + e^{-\xi T^{+}}(T^{-}
 + \pam\phi)e^{\xi T^{+}}\,, \label{cursl2}
\eea
where $\phi = \vphi T^{0}$. In the diagonal gauge
 we have
\bea
	 J_{+} &=& T^{+} - \pap\ta\,,\non \\ 
	 J_{-} &=& T^{-} - \pam\ta\,, \label{cons}
\eea
where $\ta = \vth T^{0}$. It follows from consistency
 arguments that the gauge choice for $J_{-}$ is the 
only possible choice, if we demand that $J_{-}-T^{-}
 \in \gn$. If we do not choose $J_{-} = T^{-} - 
\pam\ta$ (\ref{back}) will lead to an inconsistency.
  In order for (\ref{cursl2}) to be compatible with
 (\ref{cons}) we obtain the equations
\bea
	\pap\vphi &=&\frac{1}{\eta}\pap\eta + \eta 
\non \,, \\
	\pap\vth  &=&\pap\vphi -2\eta \,. 
\eea
The equations involving $\xi$ are similar. The equations 
can be solved to give (up to an irrelevant integration
 constant which can be set equal to 1),
 $\eta = e^{\frac{\vphi+\vth}{2}}$
 and $\xi = e^{\frac{\vphi-\vth}{2}}$. As a further
 requirement we get the following two equations connecting
 the two fields
\bea
	\pap\vphi &=& \pap\vth +
 2e^{\frac{\vphi+\vth}{2}}\,, \non \\
	\pam\vphi &=& -\pam\vth -
2e^{\frac{\vphi-\vth}{2}}\,. \label{back}
\eea
We recognize (\ref{back}) as the B\"acklund
 transformation \cite{Backlund} \cite{Curtright}
 for the Liouville model, which maps the
 Liouville field $\vphi$ onto the free field $\vth$.
 In order for the two equations in (\ref{back}) to
 be consistent we obtain
\bea
	\pap\pam\vphi + 2e^{\vphi}&=&0 \,,\non \\ 
\pap\pam\vth &=& 0\,,
\eea 
which are the equations of motion for the Liouville 
field $\vphi$ and the free field $\vth$, respectively.
 The canonical transformation  between the two fields
 implicit in (\ref{back}) is not one-to-one. For
 a discussion of this fact in the ``particle
 limit'', see ref. \cite{Bengt}. The fact that
 the mapping is not one-to-one is important and
 means that, although there exists a canonical
 transformation relating the Liouville system
 to a free system, the Liouville system is not
 trivial. The procedure outlined above can in 
principle also be used to derive relations similar
 to (\ref{back}) (B\"acklund transformations) 
for the Toda theories, although the calculations
 quickly get complex.

\setcounter{equation}{0}
\section{Discussion}
\label{disc}
We would like to mention that although we have 
only considered the generalized $A_{N}$ Toda theories 
the cases $B_{N}$ and $C_{N}$ could also be treated 
within the framework presented in this paper. In 
this paper we have exclusively dealt with the 
classical model.
It should be possible to carry out quantization 
along the usual lines. In particular it should 
be possible to calculate the central charge \cite{us}. 
However, we get seemingly infinite sums, which need 
to be properly defined. Ultimately we will face the
 same problems as in the
 Liouville theory. For a lucid discussion of the
 problems encountered in the quantization see \cite{Nicolai}.
 It should also be possible
 to study the quantum version of the ${\cal W}$ 
algebra along the lines of the finite dimensional cases.

Although we have derived the formulas in sections
 \ref{diff} and \ref{sgov} in the general case 
they apply also to the finite dimensional cases
 (ordinary Toda theories and ${\cal W}_{N}$ 
algebras); we only have to choose the parameter
 $\nu$ appropriately.

\bigskip\bigskip\noindent{\Large\bf Acknowledgement}

\bigskip\noindent The author would like to thank
 M. Vasiliev and L. Brink for enlightening discussions.

\end{document}